# Effect of Serialized Messaging on Web Services Performance

Ali Baba Dauda[#1], Zerdoumi Saber[*2], Faiz Alotaibi[+3], Muhammad A. Mustapha[#4], Muhamad Taufik Abdullah[+5]

[#]Department of Computer Engineering, University of Maiduguri, Nigeria

[*]Faculty of Computer and Information Science, University of Malaya, Malaysia

[+]Faculty of Computer and Information Science, University Putra Malaysia, Malaysia

[1, 3]ali.dauda@unimaid.edu.ng, mkurbamajnr@yahoo.com

[3,5]faiz.eid@hotmail.com, mtaufik@upm.edu.my

[2]zerdoumisaber@siswa.um.edu.my

*Abstract*— Message serialization is a format of messaging leveraging Web services to exchange data over the network. Serialized messages are processed at the server and sent as objects over the network to the client to be consumed. While, serialization process minimizes network bandwidth requirement but then incurs overhead at the communicating ends. This research contributes to the study of message exchange using HTTP across communication systems. The research identified the fundamental effect of serializing high-volume messages across network and the sources for the effects at the communication endpoints. The study utilized server - client SOAP Web services to identify the fundamental effect of serialization in the communication endpoints. SOAP messages were exchanged as XML messages over HTTP. Payload sizes (1MB – 22MB) for serialized and normal messages were exchanged through the services. The message payload, overhead, and response time were monitored and measured. The overall result indicated that is more beneficial to serialized large payload than smaller one. Generally, the serialization and deserialization cost incurred at individual ends are slightly constant irrespective of the payload size. Also, the serialization and deserialization process is insignificant to the overall transaction as the delay is below 3% of the total overhead.

*Keywords*— Client-server, Payload, Response time, Overhead,

## I. INTRODUCTION

The transfer of reliable and fast information is a basic requirement for every network communication. Serialization is one of the effective ways of transporting information across networks in form of bytes stream [1]. The serialization process converts data into object and pass over the network or store in a file. Deserialization is the reconversion of the object into its original form [2]. The principal reason for serialization is to minimize bandwidth requirement or to save space.

Web Services is one of the communication standards for messaging that binds systems on the web [3]. It exchanges SOAP messages among systems in the Web services via the transport protocol such as Hypertext Transfer Protocol (HTTP) [4]. SOAP messages exchange over the HTTP is the major standard in the server -client communication. These messages are sent in various formats to the client. Generally, the normal unformatted message transmitted via HTTP is verbose resulting to overhead. Serialized SOAP messages are exchange as objects hence minimizing bandwidth usage [5]. Although this can be advantageous but there are some tradeoffs to be judged.

This research contributes to the study of information exchange using HTTP across communication systems. The research identified the fundamental effect of serializing high-volume messages across network and the sources for the effects at the communication endpoints.

The rest of the paper is organized as follows; Section 2 provided related work on serialization and in Section 3 we described how the experiment was conducted. While Section 4 presented the discussion of the result, Section 5 provided the conclusion and future direction of the research.

## II. RELATED WORK

Previous studies were conducted to facilitate SOAP message transfer across Web services systems or applications. Serialization at the server side and deserialization at the client side were both conducted by different researchers to alleviate the cost.

The study by [6] compared binary and XML serialization in java and .net platforms using different object types and values. The study provided a depth analysis of serialization and its performance effect by measuring the time required to serialize an object. The memory utilized by each object is also taken into account. The result revealed that Java supports binary serialization better than .net. But Java is poor in handling XML, especially for deserialization.

Another study was conducted by [7] to optimize Web services performance using serialization. This was achieved by developing a reuse algorithm that differentially serializes SOAP messages. The message structures are determined and saved as templates for reuse by remote web service having similar or closely similar structure with the saved template. This process reduces the computation overhead. The only limitation is the continual growth of templates in the overall Web services.

The attempt on deserialization was conducted by [8]. In the study, the incoming message is deserialized, linked to the internal automata and matched with existing one for similarity. Only the dissimilar region of the linked is processed. This process reduces the response time, round-trip and computation to be performed at the server side. Nonetheless, the problem may tend to impede the Web services as the size of the automata grows with the message requests. Correspondingly, no procedure for garbage collection in the implementation and this can, in another way affect the performance.

Related studies [9],[10],[11],[12] investigated and improved on the performance of SOAP Web services. These studies considered and researched on the general Web services performance. Their overall finding revealed the effectiveness of response time of different message formats and file types on various web service applications.

Generally, critical analysis of how the serialization affects the respective client and server side is not well explored. This research contributes to the study of message exchange across communication systems using HTTP. The research identified the fundamental effect of serializing high-volume messages across network and the cause for the effects at the communication endpoints.

### III. METHODOLOGY

The experiment was conducted on Core i7; Duo core @ 2.30 GHz processor desktop using Weblogic 12c Server and Java 2EE platform. Two SOAP web services were implemented as server - client applications. Client request the service from the server. The server responded by sending the required service via HTTP. Normal and serialized XML messages were exchanged over HTTP as the transport protocol.

The server contains payload generator, sender and control class containing serialization method. While the client contains the receiver class with the deserialization method, service proxy class, web service port and associated object factory class to communicate with server. In both web services, the classes and the methods were annotated to be bounded during the SOAP web services communication.

When the application is executed, JAX-WS in the server creates the WSDL for the transaction and extract the endpoint (the client side linker). The WSDL exposed the web services to be available at the client side implying the service is ready to be consumed by the client.

In our case, any time the client request for service, serialized message is generated at the server endpoint and send to the client. Message size as the payload, is increased and continually pushed to the client as requested. The overall experiment was executed 50 times and transaction metrics were monitored, measured and recorded.

### IV. RESULT AND DISCUSSION

The result for the Web services transaction for both normal and serialized payloads are presented, analyzed, presented and discussed in this Section. Fig. 1, 2 and 3 show the graphs of the response time (ms) against the payload (mb).

*A. Normal data sends via HTTP*

Fig. 1. shows payload generation overhead and response time for server, client and total clinet-server transaction for normal payload.

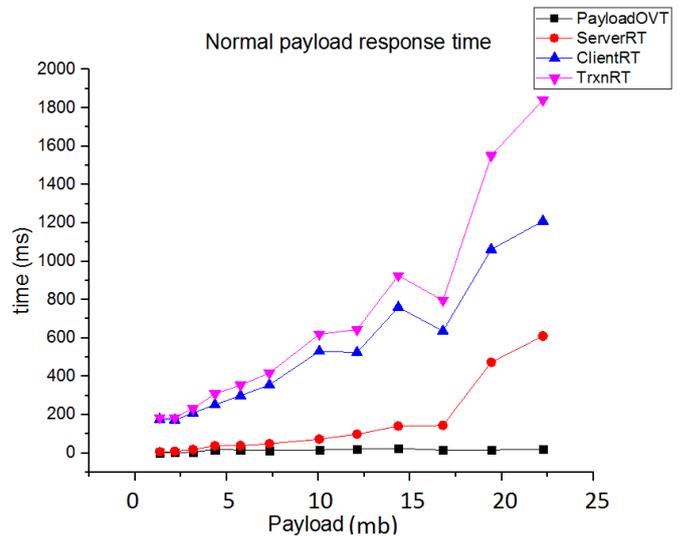

Fig. 1: SOAP over HTTP response time for normal payload transaction

As shown in fig. 1. above, the total transaction has the highest response time of 1841 ms for the highest payload of 22.2 MB. The response time picked up to maintain the normal trend as the payload size is added to the web services transaction.

The total and the client response times produced irregular trend with a similar pattern as they boomed and recessed through the transition. This continues until the payload is increased to 19.4 MB causing the total normal transaction trend to erratically change the pattern and sharply rise to 1551.3 ms and continue in this manner to the last point of payload 22.2 MB with time 1841 ms.

The average response time for both total transaction and client between payload 1.3 MB and 16.8 MB is 391.3 ms. But as the payload increased to the next level of 19.4 MB, both response times dispersed and tend to behave inconsistently.

Contrary wise to the client response time, the server response time is steady and sustained a low response time across the increasing payload. The time to respond for payloads between 1.3 MB and 16.8 MB were very low as below 150 ms. But when the payload size is increased to 19.4 MB, the response time shut up to 473.67 ms, almost 329 ms increment from the previous response time of 144.67 ms. The response time increased further with 227 ms difference from the immmediate response time (amounting to 610.67 ms) as the payload is added to 22.2 MB. Throughout the transaction, payload overhead retained stability with an almost same generation time of averagely 14.14 ms despite the successive increment in the size of the payload.

Throughout the web services transaction, the delay caused by payload generation is always the same throughout the exchange. This can be seen from the trend line in fig. 2. as it kept a parallel timing along the X-axis with minimum value of

2 ms and maximum values of 20.7 ms with a mean value of 14 ms.

The response time for the client Web services revealed to be high due to HTTP request. By and large, HTTP is a request-based protocol that perform by the requesting service. The client as the requesting service always repeatedly checks the server for any interim message; as a result, the client monopolizes the transaction thread.

The total transaction indicates to be going high with the payload as seen in the pattern in figure 3.1. This is possible due to the client response time that dominated the entire web service transaction. But contrariwise, the time at the server tends to rise only at the payload of 19.4MB. This might be due to memory swapping by the Java Virtual Machine (JVM) to allow space for an incoming message. As evidently shown in the server response trend, payload increase to 19.4 MB might be regarded as huge to the Central Processing Unit (CPU) process, therefore more space is needed to continue processing more incoming messages.

The generated payload is a sole liability of the server and is generated based on the request by the client. The trend line indicated by the payload overhead indicates the same CPU time is needed to generate and concatenate the payload to the previous payload.

*B. Serialized data sends via HTTP*

The metrics for serialized payload transaction is depicted in Fig. 2. The metrics include payload generation overhead and response time for server, client, serialization, deserialization and total clinet-server transaction for normal payload.

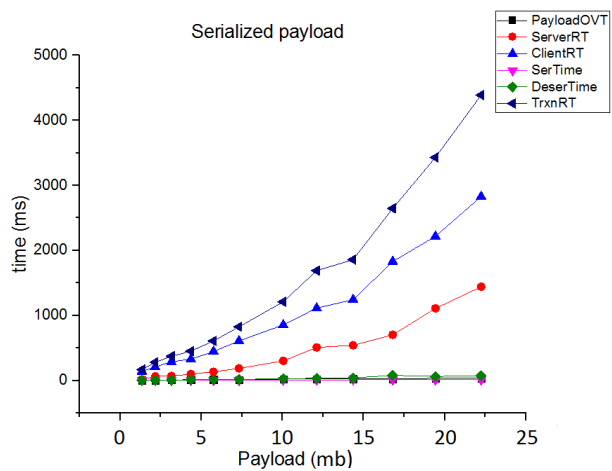

Fig. 2. SOAP over HTTP response time for serialized payload transaction

Payload generation overhead, serialization delay and deserialization delay time were steadily having almost the same relative time throughout the communication. Except for a minor deviation by the deserialization process at the client side when the load is increased to 14.4 MB. The response time dropped down and maintained the course throughout the subsequent times of the transaction. The server response time is the same as the serialization and deserialization delays at the beginning but rise when the payload is 7.3 MB at the response time of 133 ms. It then raised and deviated further and reached 1441.7 ms with the payload of 22.3 MB.

Like the normal payload transaction, the serialized payload client and total transaction time were revealed to be high. Even though the total transaction is the sum of complete delay and response time, the two has closely identical pattern during the web services transaction. Both trends maintained a perpendicular course through the transition. But at exactly 16.8 MB payload size, they both shut up high at respectively 1830 ms and 2647 ms. This pattern continued until the end of the transaction cycle with times 2830 ms and 4391 ms respectively at final payload 22.2 MB.

The response time for the client web service is almost 50% higher than the server web service response time. The client always requests services from the server and continuously checks for any incoming message from the server, thus, the client dominates most part of the transaction time. Requesting services from the server and processing the message incur process utilization at the client side. The server provides the services as functions for the transaction and tends to rise gradually with increasing payload. But the overhead at the server increase when payload reaches 19.4MB and continue to gradually grow. This might be due to memory swapping by the JVM to allow space for an incoming message. The total transaction exponentially goes high with the increasing payload shown in fig. 2. Reason for this can be attributed to the dominance of the client response time that accounts for 70% of the entire web service transaction execution time.

Serialization add reference field to the actual payload and the cumulative impact become significant during a transaction, this makes the serialization process very expensive. In addition, serialization is recursive. And when the serializing object, the descriptors as well its references are also serialized, the process tends to send more data than needed. This process causes performance overhead at the server. In the same vein, deserialization incurs overhead cost at the client side as a result of request decoding by copying from the stream and allocating new storage. Serialization and deserialization cost were insignificant in the Web services transaction as realized in fig. 2. The process accounts only 2% - 3% of the total transaction time. This indicates a low effect on the CPU at both ends.

The generated payload is a sole liability of the server and is generated based on the request by the client. The trend line by the payload overhead indicates the same CPU time is needed to generate and concatenate the payload to the previously generated one.

## C. Normal versus Serialized data

The SOAP over HTTP response times for normal and serialized payloads is showing in fig. 3. Same amount of payload is incrementally sent to the request/response process in both normal and serialized format and the results were fetched after isolating the payload generation overhead. From fig. 3, it can be clearly seen that the serialized payload has high response time compared to the normal (unserialized) payload.

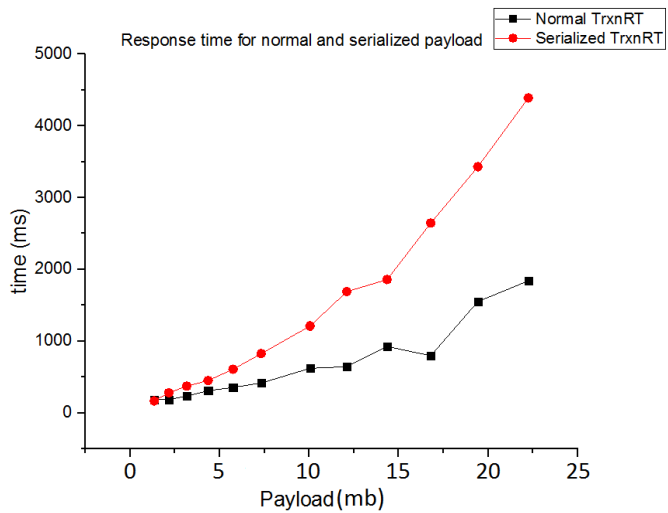

Fig. 3. SOAP over HTTP response time for normal and serialized payload

Right from the inception, the two trends do not have any common similarities. The serialized load rose and maintained perpendicular course as the magnitude of the payload is growing. The trend made a sudden increase as the payload exceeded 14.4 MB and the response time rocketed from 1860 ms to 2646 ms. This maintained a linear upward course with average increment time of 843 ms up to the end of the Web services execution cycle.

Although its initial response time (184 ms) is higher than of the serialized (166 ms), the normal payload shows an excellent inclination with a maximum response time of only 1841 ms compared to the serialized maximum response time of 4391 ms. The transition went perpendicularly as the corresponding payload is increased. It was distracted and sanked a little from 925 ms at 14.4 MB to 796 ms at 16.8 MB and suddenly surged and continue till the end of the cycle.

Unlike the serialized payload exchange, the normal payload tends to progress with little increase in the time based on the corresponding incremental payload. The normal response time dropped at 14.4 MB payload while at the same payload the serialized payload rose up.

Through the whole transaction, except in the first point, the serialized payload is higher than the normal payload. This might be as a result of initial caching in the memory. As seen in fig. 3. The serialized payload transaction time is almost 55% higher than the normal payload transaction time. The equality in the response time is attributed to the HTTP request that dominates the response time and the serialization/deserialization overheads. The verbose format of serialization and the deserialization cost a lot of memory allocation for the new incoming payload at the JVM.

## V. CONCLUSION AND FUTURE WORK

Normal message exchange does not incur overhead and have low response time as normal payload requires no conversion at any endpoint. Contrarily, serialized message exchange incurs overhead and always have a high response time. It is almost 55% higher than the normal payload response time. Serialization/deserialization cost for the serialized message is insignificant in the communication overall cost. It accounts to only 2% - 3% of the total overhead. This implies the serialization and the deserialization activities consume virtually same CPU resources and both have the same effect on the communication performance.

In both Normal and serialized message exchange communication, CPU process in the client side takes most of the communication time due to HTTP request by the client machine/application. The CPU process in the server side is low because the server only response process services when requested.

For effective performance and resource utilization at the communication endpoints, compressing the normal message might be more beneficial than serialization. Since serialization is much expensive for communication and does not take into effect the utilization of bandwidth, new techniques and protocols other than HTTP can be deployed to optimize the communication process.